\documentclass[aps,prl,reprint,twocolumn,showpacs,floatfix]{revtex4-1}
\usepackage{graphicx}
\usepackage{amsmath,amsthm,amsfonts,amssymb,amscd}

\begin{document}
\title{Quantum Speed Limit for Physical Processes}
\author{M. M. Taddei}\email{marciotaddei@if.ufrj.br}
\affiliation{Instituto de F\'isica, Universidade Federal do Rio de Janeiro, 21.941-972, Rio de Janeiro (RJ) Brazil}
\author{B. M. Escher}
\affiliation{Instituto de F\'isica, Universidade Federal do Rio de Janeiro, 21.941-972, Rio de Janeiro (RJ) Brazil}
\author{L. Davidovich}
\affiliation{Instituto de F\'isica, Universidade Federal do Rio de Janeiro, 21.941-972, Rio de Janeiro (RJ) Brazil}
\author{R. L. de Matos Filho} 
\affiliation{Instituto de F\'isica, Universidade Federal do Rio de Janeiro, 21.941-972, Rio de Janeiro (RJ) Brazil}

\begin{abstract}
The evaluation of the minimal evolution time between two distinguishable states of a system is important for assessing the maximal speed of quantum computers and communication channels. Lower bounds for this minimal time have been proposed for unitary dynamics. Here we show that it is possible to extend this concept to nonunitary processes, using an attainable lower bound that is connected to the quantum Fisher information for time estimation.  This result is used to delimit the minimal evolution time for typical noisy channels. 
\end{abstract}

\maketitle

{\it Introduction.}---Quantum mechanics imposes fundamental limits to the processing speed of any device as well as to the communication speed through any channel. Derivation of these basic limits usually assumes that such devices are noiseless, undergoing unitary evolutions \cite{MandelstamTamm,Vaidman,EberlySingh,LeubnerKiener,Fleming,Bhattacharyya,GisSabelli,BauerMello,UffHilge,Uhlmann,
Uffink,Pfeifer,PfFrohlich,AA,Anandan,HoreshMann,Pati,MargolusLevitin,Soderholm,LevitinToffoli,GiovannettiPRA,GiovannettiEPLJOB,Batle,Borras2006,
Zander,Kupferman,Frowis,Andrecut, GrayVogt,LuoZhang,Andrews,ZielinskiZych,Yurtsever,FuLiLuo,Chau,Deffner,Brody,Ashhab}. The relevant (and often unwanted) influence of the environment on processing or information-transfering systems is thus frequently ignored. On the other hand, this influence, and in particular the decoherence speed, plays an essential role in fundamental physics, especially in the understanding of the quantum-to-classical transition~\cite{Zurek}. Here we unify the description of both computation/communication speed and decoherence speed in a single framework, which deals with the maximal speed of evolution of quantum systems. 

Although much work has been done on the subject since the first major result by Mandelstam and Tamm \cite{MandelstamTamm}, scarce contributions \cite{JonesKok,CarliniBr8,Beretta,Obada,BrodyN} undertake nonunitary evolutions. In this Letter, we develop a method that allows one to derive useful saturable lower bounds for the minimal evolution time for general physical processes.  Besides recovering, in the proper limits, previous findings such as the Mandelstam-Tamm bound~\cite{MandelstamTamm}, our result allows the study of experimentally more realistic open systems and the development of a systematic approach for tackling such nonunitary evolutions. This approach relies on variational techniques, allowing one to obtain nontrivial analytical approximations to the bounds in situations where exact calculations are too involved. We exemplify the usefulness of this bound by considering typical nonunitary quantum channels.

{\it General bound for the minimal evolution time.}---We present here a general lower bound on the time $\tau$ necessary for a quantum system, evolving  under the action of some physical process, to reach a final state that has a distance $\mathcal{D}$ from its initial state.

Let $D[F_B(\hat\rho_1,\hat\rho_2)]$ be a metric on the space of quantum states that depends on $\hat\rho_1,\hat\rho_2$ solely via the Bures fidelity $F_B$,
\begin{equation}
F_B(\hat\rho_1,\hat\rho_2) := \left[{\rm tr}(\sqrt{\sqrt{\hat\rho_1} \hat\rho_2 \sqrt{\hat\rho_1}})\right]^2 .
\label{Bures}
\end{equation}
 Consider now a smooth dynamical process in this space, parametrized by $t$,  and leading to an evolution described by the density operator $\hat\rho(t)$,  such that $D[F_B(\hat\rho(t_1),\hat\rho(t_2))]$ (written as $D(t_1,t_2)$ in a shorthand notation) is a piecewise smooth function of $t_1,t_2$. 
 A bound on $D(0,\tau)$ can be obtained in terms of the integral of the quantum Fisher information for time estimation ${\cal F}_Q(t)$ along the path determined by system evolution. $\mathcal F_Q(t)$ may be defined  by $\mathcal F_Q(t) = {\rm Tr} [ \hat\rho(t)\hat{L}^2(t) ]$ \cite{BraunsteinCaves}, where the Hermitian operator $\hat{L}(t)$ is known as the symmetric logarithmic derivative operator, implicitly defined by $d\hat\rho(t)/dt = [ \hat\rho(t)\hat{L}(t) + \hat{L}(t)\hat\rho(t) ]/2 $. In order to derive the bound on $D(0,\tau)$, one applies to this metric the triangle inequality, considering a division of the interval $(0,\tau)$ into infinitesimal pieces, and one uses the relation between the Bures fidelity and the quantum Fisher information $\mathcal F_Q(t)$ \cite{BraunsteinCaves}, 
\begin{equation}
F_B (t,t+dt) = 1 - (dt)^2\mathcal F_Q(t)/4 + \mathcal O (dt)^3 .
\label{BuresFisher}
\end{equation}
This equation attaches a physical meaning to the quantum Fisher information: the square root of $\mathcal F_Q(t)$ is proportional to the instantaneous speed of separation between  two neighboring states $\hat\rho(t)$ and  $\hat\rho(t+dt)$, and can be used to delimit  the minimal statistical uncertainty in the estimation of the duration of a given physical process, as shown in  \cite{BraunsteinCaves}.

One gets then a general implicit lower bound on the evolution time $\tau$, valid for arbitrary physical processes and any metric dependent on the Bures fidelity (see  Supplemental Material  \cite{Supp}): 
\begin{equation}
\left.\sqrt{\frac{d^2 D(F_B)/d F_B^2}{2\left[d D(F_B)/d F_B\right]^3}}\right|_{F_B\rightarrow1} D(0,\tau) \leq \int_0^\tau \sqrt{\frac{\mathcal F_Q(t)}{4}}\,dt ,
\label{generalbound}
\end{equation}
where the notation $D(F_B)$ makes explicit the dependence of the metric on $F_B$. Notice that the square root on the l.h.s. of the above inequality is proportional to the curvature of $D(F_B)$  at $F_B=1$ and that this bound is invariant by a rescaling $D'=kD$. 

One should note that the r.h.s. of~\eqref{generalbound} is the Bures length, as defined by Uhlmann \cite{Uhlmann2}, of the actual path followed by the state of the system $\hat\rho(t)$.  On the other hand, it has also been shown in~\cite{Uhlmann2} that the Bures length of a geodesic joining two density operators  $\hat\rho_1$ and $\hat\rho_2$ is $ \arccos\sqrt{F_B(\hat\rho_1,\hat\rho_2)}$ \cite{arccos}, which defines a natural distance $ \mathcal{D}$ between the two states. Inserting this expression into the l.h.s. of~\eqref{generalbound}, one obtains
\begin{equation}
\mathcal D:=\arccos\sqrt{F_B\left[\hat\rho(0),\hat\rho(\tau)\right]} \leq \int_0^\tau \sqrt{\mathcal F_Q(t)/4} \ dt .
\label{boundarccos}
\end{equation}
 Since it is always possible to find a dynamical process that joins $\hat\rho(0)$ and $\hat\rho(\tau)$ along a geodesic and saturates the above inequality, one finds that  the l.h.s. of~\eqref{generalbound}, which depends only on the initial and final states, is maximized by $D(F_B) = \arccos{\sqrt{F_B}}$. Therefore, this is the optimal choice for $D(F_B)$ in~\eqref{generalbound}, and leads to an attainable bound for the minimum evolution time, valid for unitary or nonunitary processes. The above discussion makes it clear that this bound is attained if and only if the evolution occurs on a geodesic, which is, incidentally, the same condition for attainability of the Mandelstam-Tamm  bound for unitary processes with time-independent Hamiltonians~\cite{AA}.
 
Let us first consider a unitary evolution, dictated by the operator $\hat U(t)$, which leads to a simple analytical expression for the quantum Fisher information. In this case, $\mathcal F_Q(t)=4 \langle \Delta\hat H^2(t) \rangle/\hbar^2$ \cite{Boixo}, where $\langle \Delta\hat H^2(t) \rangle$ is the variance in the initial state of a Hermitian operator $\hat H(t)$ defined as 
\begin{equation}
\hat H(t):= \frac{\hbar}{i} \dfrac{d\hat U^\dagger(t)}{dt}\hat U(t) .
\label{defH}
\end{equation}
For a time-independent Hamiltonian $\hat H$, with $\hat U(t)=e^{-i\hat Ht/\hbar}$, then $\hat H(t)=\hat H$, and for $\mathcal{D}=\pi/2$ (orthogonal states), inequality~\eqref{boundarccos} leads to the Mandelstam-Tamm bound: $\tau\ge (\pi\hbar/2)/\sqrt{\langle\Delta\hat H^2\rangle}$. Eq.~\eqref{boundarccos} also recovers the known implicit bounds on $\tau$  for time-dependent Hamiltonians \cite{Uhlmann,Deffner}.

For nonunitary evolutions, bound~\eqref{boundarccos} can be hard to evaluate analytically, since the quantum Fisher information may be difficult to calculate. In these  situations, it is convenient to resort to the following purification procedure \cite{BLNR,BRL}, which allows one to rely on the simple form of $\mathcal F_Q(t)$ for unitary processes.

To each system of interest $S$, represented by the density operator $\hat\rho_S$, one assigns an environment $E$, such that the dynamics of $\hat\rho_S$ results from a unitary evolution,  corresponding to an operator $\hat U_{S,E}(t)$, of a pure state of the enlarged system $S+E$.  The quantum Fisher information  of $S+E$ is an upper bound to the quantum Fisher information of system $S$, since, from the point of view of parameter-estimation theory,  $S+E$ does not contain less information about the parameter $t$ than $S$ alone.  There are, in fact, infinitely many different evolutions of $S+E$ corresponding to the same evolution of system $S$, each of those leading to a possibly different value of the quantum Fisher information $\mathcal{C}_Q(t)$ of $S+E$. This freedom is integrally expressed by writing the purified unitary evolution in $S+E$ as $\hat u_E(t)\hat U_{S,E}(t)$, where $\hat u_E(t)$ is any unitary operator acting only on $E$. Defining $\hat{\mathcal{H}}_{S,E}(t)$  by inserting the evolution operator $\hat u_E(t)\hat U_{S,E}(t)$ into~\eqref{defH}, one can write $\mathcal C_Q(t)=4\langle\Delta\hat{\mathcal{H}}_{S,E}^2(t)\rangle/\hbar^2$.  Then, for any  upper bound $\mathcal{C}_Q(t)$ to  $\mathcal{F}_Q(t)$, one can obtain an implicit lower bound to the evolution time $\tau$, given by 
\begin{equation}
 \mathcal D \leq \int_0^\tau \sqrt{\mathcal C_Q(t)/4} \ dt = \int_0^\tau \sqrt{\langle\Delta\hat{\mathcal{H}}_{S,E}^2(t)\rangle}/\hbar \, dt.
\label{boundarccoscq}
\end{equation} 
Since $\mathcal{C}_Q(t)$ can be straightforwardly evaluated, the above bound may be easier to handle than bound~\eqref{boundarccos}.    
However, it can only be tight when $\mathcal{C}_Q(t)=\mathcal{F}_Q(t)$, in which case it reduces to bound~\eqref{boundarccos}.  
In fact, as it was shown in \cite{BLNR,BRL}, it is always possible to fulfill this condition by minimizing $\mathcal{C}_Q(t)$ over all operators  $\hat u_E(t)$, for given  $\hat U_{S,E}(t)$.
As  $\langle\Delta\hat{\mathcal{H}}_{S,E}^2(t)\rangle$  only depends on $\hat u_E(t)$ through $\hat h_E(t)$,
\begin{equation}
\hat h_E(t) := \frac\hbar i \dfrac{d\hat u_E^\dagger(t)}{dt}\hat u_E(t) ,
\label{defhE}
\end{equation}
the minimization can be performed with respect to $\hat h_E(t)$~\cite{BLNR}. Notice that in some practical situations, it can be advantageous to restrict the set of operators $\hat h_E(t)$ over which the optimization is done in order to obtain more tractable, albeit still useful  bounds on $\tau$. In the following, we present examples that illustrate the power and usefulness of this approach. 


{\it Amplitude-damping channel.}---Let $S$ be a two-state system (states $\{|0\rangle,|1\rangle\}$), and $E$ its environment, which is chosen to start in state $|0\rangle_E$. The amplitude-damping channel is described by the map
\begin{subequations}\begin{align}
|0\rangle|0\rangle_E &\rightarrow |0\rangle|0\rangle_E \label{amp1}\,,\\
|1\rangle|0\rangle_E &\rightarrow  \sqrt{P(t)} |1\rangle|0\rangle_E +\sqrt{1-P(t)} |0\rangle|1\rangle_E , \label{amp2}
\end{align}\label{amplitudechannel}
\end{subequations}
where the state $|1\rangle_E$ is orthogonal to $|0\rangle_E$, and 
the time-dependence of $P(t)$ reflects the damping dynamics. We consider here the paradigmatic exponential decay, with rate $\gamma$, so that $P(t)=e^{-\gamma t}$. We note that, for the above map, the environment may also be considered as a qubit. This channel, which corresponds to a \textit{nonunitary} evolution of $S$, can be described by the unitary evolution operator acting on $S+E$,
\begin{equation}
\hat U_{S,E}(t)=\exp[-i\Theta(t)(\hat\sigma_+\hat\sigma_-^{(E)}+\hat\sigma_-\hat\sigma_+^{(E)} )] ,
\label{Uamplitude}
\end{equation}
where $\hat\sigma_\pm$ and $\hat\sigma_\pm^{(E)}$ are raising and lowering operators acting respectively on the system and environment qubits, and $\Theta(t)=\arccos\sqrt{P(t)}$.

Setting $\hat u_E(t)$ as an identity operator and inserting the variance  of $\hat{\mathcal H}_{S,E}(t)$, obtained from the unitary operator~\eqref{Uamplitude} via definition~\eqref{defH}, into~\eqref{boundarccoscq}, it is straightforward to show that $\tau$ is bounded by
\begin{equation}
\gamma\tau \geq 2 \ln\sec ( \mathcal D / \sqrt{ \langle\hat\sigma_+\hat\sigma_-\rangle } ) .
\label{resultdecay}
\end{equation}
Notice that, for the above process, the distance of the evolved state from the initial state can reach at most $\langle \hat\sigma_+ \hat\sigma_-\rangle \pi/2$.
Bound~\eqref{resultdecay} saturates either for $S$ initially in the ground state ($\langle  \hat\sigma_+ \hat\sigma_-\rangle =0$), when the system does not evolve at all, or for $S$ initially in the excited state ($\langle  \hat\sigma_+ \hat\sigma_-\rangle =1$), 
meaning that we have already chosen the purified evolution that yields $\mathcal C_Q(t)=\mathcal F_Q(t)$ for these situations. Furthermore, the fact that the bound is saturated implies that the amplitude-damping channel connects the states $|1\rangle$ and $|0\rangle$ 
along a geodesic path, which includes mixed states. This remains valid for any monotonically decreasing $P(t)$, with $P(0)=1$,  since, under this condition, changing the form of $\Theta(t)$ in~\eqref{Uamplitude} corresponds to a mere rescaling of the time parameter, which does not change the path in state space followed by a given initial state. 

{\it Markovian dephasing.}---System $S$ is again a single qubit whose nonunitary evolution is described by a map that makes use, as before, of an ancilla qubit starting in state $|0\rangle_E$,
\begin{equation}\begin{split}
|0\rangle|0\rangle_E &\rightarrow e^{-i\omega_0t/2} \left( \sqrt{P(t)} |0\rangle|0\rangle_E + \sqrt{1-P(t)} |0\rangle|1\rangle_E \right) ,\\
|1\rangle|0\rangle_E &\rightarrow e^{ i\omega_0t/2} \left( \sqrt{P(t)} |1\rangle|0\rangle_E - \sqrt{1-P(t)} |1\rangle|1\rangle_E \right) ,
\end{split}\label{dephasingchannel}
\end{equation} where $\hbar\omega_0$ is the energy difference between the qubit levels, $P(t):=(1+e^{-\gamma t})/2$, and $\gamma$ is the phase-decay constant. The corresponding evolution operator is
\begin{equation}
\hat U_{S,E}(t) = e^{-i\omega_0 t \hat Z/2} e^{-i\arccos\sqrt{P(\gamma t)} \hat Z \hat Y^{(E)}} ,
\label{evol}
\end{equation}
where $\hat Z$ and $\hat Y^{(E)}$ are Pauli operators acting on the system and on the environment qubits, respectively. 

\begin{figure}[tb]
\includegraphics[width=\columnwidth]{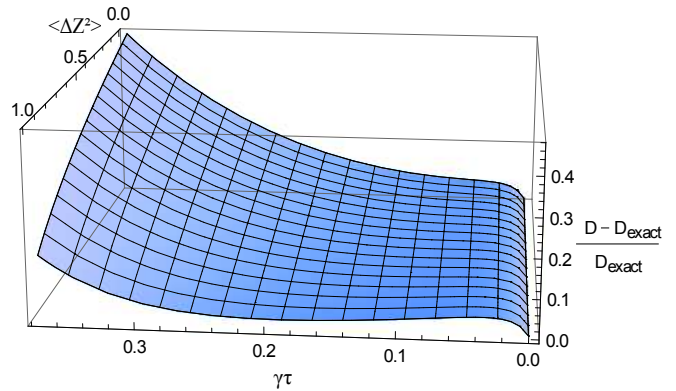} 
\caption{Relative difference between bound on $\mathcal D$ for single-qubit dephasing~\eqref{MTbound1qbit} and the exact result,  as a function of the dimensionless time $\gamma\tau$, for different values of  $\langle\Delta\hat Z^2 \rangle$ (different initial states), with $r=8$.}
\label{1qbit}%
\end{figure}

In order to find the best possible bound on $\tau$  within our approach, we now minimize $\mathcal{C}_Q(t)$  over the whole set of Hermitian $2\times 2$ operators  $\hat h_E(t)$. 
The minimum, written in terms of the (constant) variance of $\hat Z$, is~\cite{Supp}
\begin{equation}
\mathcal{C}_Q^{\rm opt}(t) = \langle\Delta\hat Z^2 \rangle\left[ \omega_0^2e^{-2\gamma t} + \gamma^2(e^{2\gamma t}-1)^{-1} \right]  ,
\label{deltaHmin1}
\end{equation}
which reduces to $\mathcal F_Q(t)$ for pure initial states, in which case bound~\eqref{boundarccoscq} reduces to~\eqref{boundarccos}. 
 The above equation leads to an implicit  bound on $\tau$, given, in terms of elliptic integrals of the second kind $E(y,k)$, by
\begin{align}
& \mathcal D \leq  \tfrac{1}{2} \sqrt{\langle\Delta\hat Z^2 \rangle} \sqrt{r^2+1} \label{MTbound1qbit} \\
\nonumber &		\times   \left[ E\left(\frac{\pi}{2},\frac{r}{\sqrt{r^2+1}}\right) 	- E\left(\arcsin e^{-\gamma\tau}, \frac{r}{\sqrt{r^2+1}}\right) \right] ,
\end{align}
with $r:=\omega_0/\gamma$. Eq.~\eqref{MTbound1qbit} consistently guarantees the eigenstates of $\hat Z$ not to evolve. This bound  is compared to an exact calculation of $\mathcal D$ in Fig.~\ref{1qbit}, which shows that it stays close to the exact result up to the first minimum of the latter. Another feature of~\eqref{MTbound1qbit} is that it captures the fact that the evolved and initial states never become orthogonal for $r$ under a critical value $r_{\rm crit} \simeq 2.6$~\cite{Supp}.

In the extreme cases  $\gamma\rightarrow0$ and $\omega_0\rightarrow0$, inequality~\eqref{MTbound1qbit} yields simple analytical expressions for the bound on the minimal time. For the former, the Mandelstam-Tamm bound~\cite{Vaidman,Bhattacharyya} is recovered, since the process becomes unitary; and for the latter
\begin{align}
     \gamma \tau       \geq  \ln \sec \left(2 \mathcal D /\sqrt{\langle\Delta\hat Z^2 \rangle}\right) ,
\label{1qbit-r=0}
\end{align}
which saturates for pure initial states with $\langle\Delta\hat Z^2 \rangle=1$. In this situation, pure Markovian dephasing ($\omega_0\rightarrow0$) links a pure state to a fully mixed state through a geodesic path in state space. Notice that, for the above process, the distance of the evolved state from the initial state can reach at most $\sqrt{\langle\Delta\hat Z^2 \rangle}\,\pi/4$.

{\it Minimum evolution time and entanglement.}---We now investigate the effect of subsystem correlations on the evolution speed of a compound system. We consider the Markovian dephasing of an N-qubit system where each qubit interacts only with its own environment, as described by~\eqref{dephasingchannel}, and compare how different initial-state correlations (possibly entanglement) affect the evolution speed.
The evolution operator is $\hat u_E(t) \hat U_{S,E}(t)$, with
\begin{equation}
\hat U_{S,E}(t) =  \prod_{i=1}^N e^{-i\omega_0 t \hat Z_i/2} e^{ - i\arccos\sqrt{P(\gamma t)} \hat Z_i\hat Y^{(E)}_i } ,
\label{evolN}
\end{equation}
where $\hat Z_i$ ($\hat Y_i^{(E)}$) is a Pauli operator acting on the $i$-th system (environment) qubit.  Since $\hat h_E(t)$  now belongs to a $2^N\times2^N$ space, the full minimization of $\mathcal C_Q(t)$  is rather cumbersome for large values of $N$. Hinging on the symmetry of the system, we resort instead to minimization over a three-parameter family of Hermitian operators:
\begin{equation}
\hat h_E(t) =  \sum_{i=1}^N \left[\alpha(t) \hat X^{(E)}_i + \beta(t)  \hat Y^{(E)}_i + \delta(t) \hat Z^{(E)}_i\right] ,
\label{hE1N}
\end{equation}
where $\alpha(t)$, $\beta(t)$, and $\delta (t)$ are optimization variables. We get then~\cite{Supp}
\begin{align}
\mathcal C_Q^{\rm opt}(t) & =  \langle\Delta\mathcal{\hat Z}^2\rangle \left[ \dfrac{\omega_0^2 N^2}{Nq(e^{2\gamma t}-1) + 1} + 
\frac{\gamma^2N/q}{e^{2\gamma t}-1}\right],  \label{deltaHminNq}
\end{align}
where 
$q:=\langle\Delta \hat{\mathcal{Z}}^2\rangle/(1-\langle\hat{\mathcal{Z}}\rangle^2)$, $\hat{\mathcal{Z}}=\sum_j\hat Z_j/N$, and the averages are taken in the initial state.
We note that $0\leq q \leq 1$;  for a separable state, $q\leq1/N$ (equality if symmetrical on the N qubits). The values $q=1$ and $\langle\Delta\hat{\mathcal{Z}}^2\rangle=1$, achievable for GHZ states $[|0\dots0\rangle+e^{i\phi}|1\dots1\rangle]/\sqrt{2}$, yield a lower bound valid for any initial state. For $q=1$, the bound on the minimum evolution time scales as $\tau\sim 1/N$ throughout, a prediction validated by exact calculations with GHZ states, see Supplemental Material \cite{Supp}.

\begin{figure}[tb]
\includegraphics[width=\columnwidth]{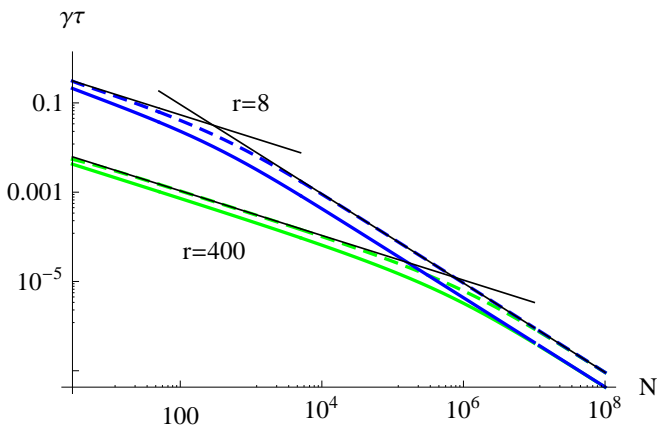}
\caption{(Color online) Lower bound (solid curves) on time for separable, symmetric state with $\langle{\Delta \hat{\mathcal Z}^2}\rangle=1$ to reach $\mathcal D \simeq 94\%$ of the maximal distance (${F_B}=1\%$), measured in dimensionless units $\gamma\tau$ as a function of number of qubits $N$, calculated numerically from~\eqref{deltaHminNq}. Results from exact calculations~\cite{Supp} are plotted for comparison (dashed curves). For the black (blue) curves, $r=8$, for the light gray (green) curves, $r=400$; the asymptotes are proportional to $1/N$ and $1/\sqrt N$.}
\label{figtauNtrans}
\end{figure}

For separable states, on the other hand, the lower bound goes from a $\tau \sim 1/\sqrt N$ dependence for $\gamma\sqrt{N}\ll\omega_0$ to $\sim 1/N$  for $\gamma\sqrt{N}\gg\omega_0$, as shown in Fig.~\ref{figtauNtrans}. This transition to faster behavior can be corroborated via direct calculations on symmetric, separable states \cite{Supp}.

This is a striking result, clearly distinct from the one corresponding to unitary evolution. It has already been seen in the literature \cite{GiovannettiEPLJOB,Batle,Borras2006,Zander,Kupferman} that, for unitary processes, entanglement is a resource that enhances the speed of evolution, so that the separation time improves from a $\tau\sim1/\sqrt N$ scaling (separable, slow state) to $\tau\sim1/N$ (entangled, fast state). However, for  the nonunitary evolution here considered,  the minimum evolution time for separable states, while scaling with $1/\sqrt{N}$ for small $N$, eventually scales as $1/N$  for $\gamma\sqrt{N}\gg \omega_0$, no matter how small is the dephasing rate. Under this condition, the evolution speeds of separable and entangled states scale in the same way with respect to the number of qubits.

{\it Conclusion.}---We have derived  an attainable lower bound for the minimal evolution time of dynamical systems through a geometrical approach.  This bound applies to both unitary and nonunitary processes, and is obtained by comparing the actual path followed by the system in state space with the distance between the initial and final states along a geodesic path, defined by a metric that is expressed in terms of the Bures fidelity.  Whenever the evolution between two states is along this geodesic, the bound is tight. Furthermore, it encompasses several special cases discussed in the literature, including unitary evolutions and mixed initial states. 

This bound, expressed by Eq.~\eqref{boundarccos},  yields the proper speed limit for general physical processes, and reduces,  for unitary processes with time-independent Hamiltonians, to the Mandelstam-Tamm bound. This result invalidates claims that there is no general bound valid for all possible (unitary and nonunitary) quantum evolutions \cite{JonesKok}.  It is important to note that our  general bound depends on the quantum Fisher information of the system, rather than the initial variance of the Hamiltonian of the system alone. 

For situations when the general bound is too hard to evaluate, we have  introduced a more tractable bound, based on a purification procedure that leads to attainable upper bounds for the quantum Fisher information. 


The usefulness of this bound is exemplified by considering typical nonunitary quantum channels. For the amplitude channel, it leads to a tight bound, which evidences that the evolution between the initial and final orthogonal pure states is along a geodesic path though mixed states. For a dephasing channel, it yields very good lower bounds for the minimal evolution time between two non-orthogonal states. For $N$-qubit dephasing, the evolution speed-up due to entanglement of its subsystems, previously demonstrated for unitary evolution,  is shown to hold, in the nonunitary case, also for separable states. 

Our general result allows the estimation of the impact of the environment on the speed of quantum computation and information processing. It is also relevant for the estimation of thermalization and decoherence times. 

The authors acknowledge the support of the Brazilian agencies CNPq, CAPES, FAPERJ, and the National Institute of Science and Technology for Quantum Information.

\newpage
\onecolumngrid
\appendix
			\setcounter{equation}{0}
			\renewcommand{\theequation}{S\arabic{equation}}
			\setcounter{figure}{0}
			\renewcommand{\thefigure}{S\arabic{figure}}

{\LARGE \center Supplemental material}

\section{A. Geometric derivation of the general bound}
Let $D(F_B)$ be any function dependent only on the Bures fidelity $F_B$ defined in \eqref{Bures}. We denote by $F_B(t_1,t_2)$ the fidelity between the states of the same system at times $t_1$, $t_2$, i.e. a shorthand notation for $F_B\left[\hat\rho(t_1),\hat\rho(t_2)\right]$; $D(t_1,t_2)$ 
will analogously be a shorthand notation for $D\{F_B\left[\hat\rho(t_1),\hat\rho(t_2)\right]\}$. Here we consider dynamical evolutions of the system and functions $D(F_{B})$ so that $D(t_1,t_2)$ can be considered as a piecewise smooth metric on the space of quantum systems, which implies that $\partial D(t,x)/\partial x$ is different from zero when $x\rightarrow t$.  Dividing a given time interval into an arbitrary number of smaller ones, and using the triangle inequality for distances, one gets
\begin{equation}
D(0,\tau) \leq \sum_{i=1}^n D[(i-1)\Delta t,i\Delta t] ,
\label{triangledisc}
\end{equation}
where $\Delta t=\tau/n$. An expansion in the second argument of the $i$-th term of the sum around $(i-1)\Delta t$ yields
\begin{equation}
D[(i-1)\Delta t,i\Delta t] = \left. \frac{\partial D[(i-1)\Delta t,x]}{\partial x}\right|_{x\rightarrow(i-1)\Delta t}\Delta t + \mathcal O(\Delta t^2) ,
\label{exp}
\end{equation}
so that, when $n\rightarrow\infty$, one finds
\begin{equation}
D(0,\tau) \leq \int_0^\tau \left.\dfrac{\partial D[t,x]}{\partial x}\right|_{x\rightarrow t} dt .
\label{triangle}
\end{equation}
The integrand can be calculated using the chain rule,
\begin{equation}
\left.\dfrac{\partial D(t,x)}{\partial x}\right|_{x\rightarrow t} = \left[\dfrac{d D(F_B)}{d F_B} \dfrac{\partial F_B(t,x)}{\partial x}\right]_{x\rightarrow t} .
\label{chain}
\end{equation}
Using Eq.~\eqref{BuresFisher} of the main text, we see that $\partial F_B(t,x)/\partial x$ tends to zero when $x\rightarrow t$. Since $D(t,x)$ is a piecewise smooth function of $x$, $\left.\partial D(t,x)/\partial x\right|_{x\rightarrow t}$ is diferent from zero, except for the trivial case of a non-evolving state, and hence $d D(F_B)/d F_B$ must diverge in the $x\rightarrow t$ limit. This indeterminacy can be removed with l'H\^opital's rule, 
\begin{equation}
\left.\dfrac{\partial D(t,x)}{\partial x}\right|_{x\rightarrow t}= \left.\dfrac{\dfrac{\partial F_B(t,x)}{\partial x} }{\dfrac{1}{d D(F_B)/dF_B}}\right|_{x\rightarrow t}
= \left. \dfrac{ \dfrac{\partial^2 F_B(t,x)}{\partial x^2} } 
											{\dfrac{d}{dF_B}\left[\dfrac{1}{d D(F_B)/d F_B}\right]\dfrac{\partial F_B(t,x)}{\partial x}} \right|_{x\rightarrow t} .
\label{l'H}
\end{equation}
While the numerator is proportional to $\mathcal F_Q(t)$ due to \eqref{BuresFisher}, the denominator can be written, by multiplying and dividing by $dD/dF_B$, as
\begin{equation}
\left. \left( -\dfrac{1}{\left[d D(F_B)/d F_B\right]^3} \dfrac{d^2 D(F_B)}{d F_B^2}\right) 
																								\left( \dfrac{dD(F_B)}{dF_B} \dfrac{\partial F_B(t,x)}{\partial x}\right)  \right|_{x\rightarrow t} .
\label{denominator}
\end{equation}
The first factor in parentheses can be calculated independently of $t,x$ by replacing the limit $x\rightarrow t$ by $F_B\rightarrow1$ since the metric $D$ only depends on $t,x$ through $F_B(t,x)$. This first factor is actually proportional to the curvature of the curve $D(F_{B})$ at $F_{B}=1$. The second factor in parentheses is simply a recurrence of $\left.\partial D(t,x)/\partial x\right|_{x\rightarrow t}$, the term we are calculating. Substituting in \eqref{l'H} and rearranging the terms, one finds
\begin{equation}
\left[\dfrac{\partial D(t,x)}{\partial x}\right]_{x\rightarrow t}^2 = 
\left. \dfrac{2\left[\dfrac{d D(F_B)}{d F_B}\right]^3}{\dfrac{d^2 D(F_B)}{d F_B^2}} \right|_{F_B\rightarrow1} \dfrac{\mathcal F_Q(t)}{4} .
\label{dDdx'}
\end{equation}
Taking the square root of the r.h.s. of the above equation into \eqref{triangle}, and rearranging the terms, one obtains the general bound \eqref{generalbound}, valid for any piecewise smooth metric $D[F_B(t,x)]$ depending only on the Bures fidelity $F_B(t,x)$. One should note that \eqref{dDdx'} also holds for non-evolving states.

\section{B. Dephasing channel: derivation of bound}
Defining
\begin{equation}
\hat H_{S,E}(t) := \frac{\hbar}{i} \frac{d\hat U_{S,E}^\dagger(t)}{dt}\hat U_{S,E}(t),
\label{defHSE}
\end{equation}
it is straightforward to show that
\begin{equation}
\hat{\mathcal{H}}_{S,E}(t) = \hat H_{S,E}(t) + \hat{h'}(t) ,
\label{mathcalH}
\end{equation}
where $\hat{h'}(t):= \hat U_{S,E}^\dagger(t) \hat{h}_E(t) \hat U_{S,E}(t)$. The variance $\langle\Delta\hat{\mathcal H}_{S,E}^2(t)\rangle$ can then be cast in the form
\begin{equation}
\langle \Delta\hat{\mathcal H}_{S,E}^2(t) \rangle = \langle \Delta\hat H_{S,E}^2(t) \rangle + \langle \Delta\hat h_{E}^2(t) \rangle 
+ 2 \Re [ \langle \hat{h'}(t) \hat H_{S,E}(t) \rangle - \langle\hat{h'}(t)\rangle \langle\hat H_{S,E}(t) \rangle ] .
\label{DeltamathcalH}
\end{equation}

Let $|\psi_0\rangle|0\rangle_E$ be the initial state of $S+E$. The N-qubit dephasing is described by \eqref{evolN}, from which one gets
\begin{align} 
\hat H_{S,E}(t) & = \hbar\omega_0 N \hat{\mathcal Z}/2  + \frac{\hbar\gamma/2}{\sqrt{e^{2\gamma t}-1}} \sum_{i=1}^N\hat Z_i \hat Y_i^{(E)} ,
\label{HcomN} \\
\langle \hat H_{S,E}(t) \rangle & = \hbar\omega_0N \langle \hat{\mathcal Z} \rangle /2 , \label{medH}
\end{align}
which imply that 
\begin{equation}
\langle \Delta\hat H_{S,E}^2(t) \rangle = N^2 \frac{\hbar^2\omega_0^2}{4} \langle\Delta \hat{\mathcal{Z}}^2\rangle + \frac{N\hbar^2\gamma^2/4}{e^{2\gamma t}-1} .
\label{medDeltaH2}
\end{equation}
From \eqref{hE1N},
\begin{equation}
\langle\hat{h'}(t)\rangle = \delta(t)N\left(2P(t)-1\right) - \alpha(t) N \langle \hat{\mathcal Z}\rangle 2\sqrt{P(t)}\sqrt{1-P(t)} ,
\label{h'med}
\end{equation}
where $P(t):=(1+e^{-\gamma t})/2$, and
\begin{equation}\begin{split}
\langle\Delta\hat{h'}^2(t)\rangle = & N [\alpha^2(t)+\beta^2(t)+\delta^2(t)]
																	+ \alpha^2(t) \left( N^2\langle\Delta\hat{\mathcal{Z}}^2\rangle - N \right) 4P(t)\left[1-P(t)\right]  \\
 &- N \delta^2(t) \left[2P(t)-1\right]^2 + 2 \alpha(t) \delta(t) N \langle\hat{\mathcal Z}\rangle 2\sqrt{P(t)}\sqrt{1-P(t)}\left[2P(t)-1\right] .
\end{split}\label{medDeltah2}
\end{equation}
From the previous equations,
\begin{equation}
2 \Re [ \langle \hat{h'}(t) \hat H_{S,E}(t) \rangle - \langle\hat{h'}(t)\rangle \langle\hat H_{S,E}(t) \rangle ] = 
- 2\alpha(t) \hbar\omega_0 N^2\langle\Delta\hat{\mathcal{Z}}^2\rangle \sqrt{P(t)}\sqrt{1-P(t)} + \frac{N\hbar\gamma\beta(t)}{\sqrt{e^{2\gamma t}-1}} \langle \hat{\mathcal Z}\rangle , 
\label{corrhH}
\end{equation}
and $\langle\Delta\hat{\mathcal H}_{S,E}^2(t)\rangle$, according to \eqref{DeltamathcalH}, is the sum of \eqref{medDeltaH2}, \eqref{medDeltah2}, and \eqref{corrhH}.

One now minimizes over $\alpha(t)$, $\beta(t)$, $\delta(t)$ for each $t$. From $\partial \langle\Delta\hat{\mathcal H}_{S,E}^2(t)\rangle / \partial \beta = 0$, one gets
\begin{equation}
\beta(t)= - \frac{\hbar\gamma}{2\sqrt{e^{2\gamma t}-1}} \langle \hat{\mathcal Z}\rangle .
\label{beta}
\end{equation}
Conditions $\partial \langle\Delta\hat{\mathcal H}_{S,E}^2(t)\rangle / \partial \alpha = 0$ and $\partial \langle\Delta\hat{\mathcal H}_{S,E}^2(t)\rangle / \partial \delta = 0$ lead to a linear system of equations, with solutions
\begin{align}
\alpha(t)= {} & \dfrac{\hbar\omega_0 e^{\gamma t} \sqrt{e^{2\gamma t}-1} N q / 2}	{1 +(e^{2\gamma t}-1)N q } ,
\label{alpha} \\
\delta(t)= {} & -\dfrac{\hbar\omega_0 e^{2\gamma t} \langle\hat{\mathcal Z}\rangle N q / 2}	{ 1 + (e^{2\gamma t}-1)N q } ,
\label{delta}
\end{align}
where $q:=\langle\Delta\hat{\mathcal{Z}}^2\rangle/(1 - \langle \hat{\mathcal Z}\rangle^2)$. By replacing \eqref{beta}, \eqref{alpha}, and \eqref{delta} into  \eqref{medDeltaH2}, \eqref{medDeltah2}, and \eqref{corrhH}, one finds
\begin{equation}
4\langle\Delta\hat{\mathcal H}_{S,E}^2(t)\rangle = \hbar^2 \langle\Delta\mathcal{\hat Z}^2\rangle \left\{ \dfrac{\omega_0^2 N^2 }{Nq(e^{2\gamma t}-1) + 1} +  \frac{\gamma^2 N /q}{e^{2\gamma t}-1} \right\},
\label{resultN}
\end{equation}
and \eqref{deltaHminNq} follows because $\mathcal C_Q(t)=4\langle\Delta\hat{\mathcal{H}}_{S,E}^2(t)\rangle/\hbar^2$. The single-qubit result \eqref{deltaHmin1} is recovered by taking $N=1$ (notice that $q=1$ in this case).

\section{C. Dephasing channel: exclusion window for single-qubit case}
\begin{figure}[tb]\centering
\includegraphics[width=.4\columnwidth]{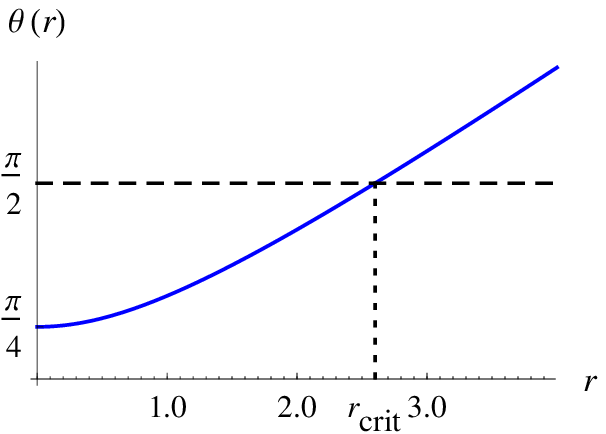}
\caption{Bound $\theta(r)$ for $\mathcal D$ corresponding to single-qubit Markovian dephasing, given by the right-hand side of \eqref{MTbound1qbit} for evolution time $\tau\rightarrow\infty$ and $\langle \hat Z\rangle=0$. Values under $\pi/2$ indicate regions where the bound guarantees that no state turns orthogonal to itself.} \label{arc-r-infinito}
\end{figure}

By taking in the r.h.s. of \eqref{MTbound1qbit} the limit $\tau \rightarrow \infty$, with $\langle\Delta\hat Z^2\rangle=1$, and plotting as a function of $r$, we obtain Fig.\ref{arc-r-infinito}.
There clearly is a region $r<r_{\rm crit}$ where $\mathcal D$ does not reach $\pi/2$ for any finite $\tau$, hence the evolved and initial states never become orthogonal (for any initial state). This exclusion window -- calculations give $r_{\rm crit} \simeq 2.6$ -- yields a simple criterion for defining the regimes of strong ($r>r_{\rm crit}$) and weak ($r<r_{\rm crit}$) dephasing;  exclusion windows for reaching given, shorter distances can be calculated in an analogous way.

\section{D. Dephasing channel: $N$-dependence of the most general bound}
\begin{figure}[tb]\centering
\includegraphics[width=.3\columnwidth]{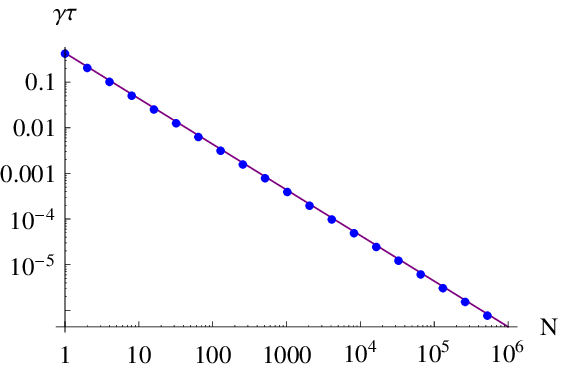} \includegraphics[width=.3\columnwidth]{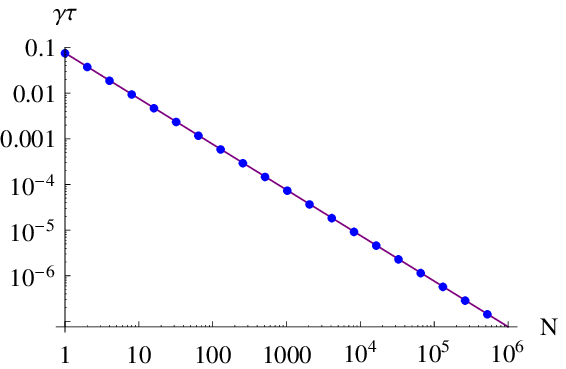} \includegraphics[width=.3\columnwidth]{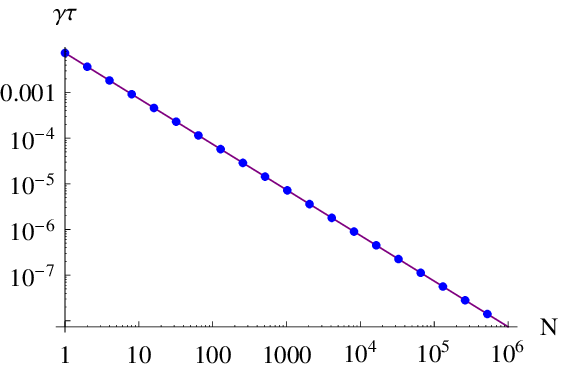} \caption{Most general lower bound on time $\gamma\tau$ for state to reach $\mathcal D=94\%$ of the maximal distance (${F_B}=1\%$), calculated numerically from \eqref{boundGHZ}, as a function of $N$, with $r=8$, $40$ and $400$, respectively (error bars smaller than symbols). The straight line, proportional to  $1/N$, obeys \eqref{boundGHZcoeff}.}%
\label{tauNGHZ}
\end{figure}

The most general bound for $N$ qubits under Markovian dephasing, attained with parameters corresponding to the GHZ state, is
\begin{equation}
2\mathcal D \leq \sqrt{N} \int_0^{\gamma\tau} \sqrt{ r^2 \dfrac{N}{N(e^{2u}-1) + 1} +  \frac{1}{e^{2u}-1} } du .
\label{boundGHZ}
\end{equation}
An equally general, albeit slightly larger, upper bound for $\mathcal D$ leads to an analytical expression for the bound on $\tau$. It is found by considering $N(e^{2\gamma\tau}-1) \gg 1$. One gets then
\begin{equation}
 \sqrt{N} \sqrt{r^2 + 1} \arctan \sqrt{ e^{2\gamma\tau}-1 }\ge 2 \mathcal D ,
\label{boundGHZmed}
\end{equation}
which yields, for $N\gg1$,
\begin{equation}
\tau \geq \frac{1}{N} \dfrac{4\mathcal D^2}{\gamma(r^2 + 1)} .
\label{boundGHZmedfin}
\end{equation}
This expression explicitly exhibits a $\tau\sim1/N$ dependence. An alternative estimate, which leads to a better approximation of the integral in \eqref{boundGHZ}, is found in the $r\gg1$ limit, yielding, to lowest order in $1/N$:
\begin{equation}
\tau \geq \frac{1}{N} \frac{2\mathcal D}{\omega_0} \left( 1 + \frac{\mathcal D}{r} \right) .
\label{boundGHZcoeff}
\end{equation}
Fig.~\ref{tauNGHZ} displays the numerically calculated bound, which is compared to the approximation \eqref{boundGHZcoeff} .

\section{E. Dephasing channel: $N$-dependence of the bound for separable, symmetric states}
From \eqref{deltaHminNq}, the bound for separable states can  be written in terms of elliptic functions of the second kind $E(y,k)$,
\begin{equation}
2 \mathcal D \leq  \sqrt{1 - \langle \hat Z \rangle^2} \sqrt{N} \sqrt{r^2+1} 
	  \left[ E\left(\frac{\pi}{2},\frac{r}{\sqrt{r^2+1}}\right) 	- E\left(\arcsin e^{-\gamma\tau}, \frac{r}{\sqrt{r^2+1}}\right) \right] .
\label{boundSep}
\end{equation}
An expansion to lowest order on $\gamma\tau$ yields $\tau\sim1/N$:
\begin{equation}
 \tau \geq \dfrac{1}{N} \dfrac{ 2 \mathcal D^2 } {\gamma  \langle \Delta\hat Z^2 \rangle} .
\label{boundSep-u}
\end{equation}
For $r\gg1$,  one gets instead, from \eqref{boundSep},  
\begin{equation}
2 \mathcal D \leq \sqrt{\langle \Delta\hat Z^2 \rangle} \sqrt{N} r (1-e^{-\gamma\tau}) ,
\label{boundSep-r}
\end{equation}
which leads to a  $\tau\sim1/\sqrt{N}$ dependence,
\begin{equation}
\tau \geq \dfrac{1}{\sqrt N } \dfrac{2\mathcal D}  { \omega_0 \sqrt{\langle \Delta\hat Z^2 \rangle} } .
\label{boundSep-rtauN}
\end{equation}

An estimate of the transition between \eqref{boundSep-u} and \eqref{boundSep-rtauN} is obtained by equating the two respective contributions to the bound on $\mathcal D$. This leads to
\begin{align}
\gamma\tau_{\rm tr} \simeq 2/r^2 \label{gammatautrans} \\
\sqrt{N_{\rm tr}} \simeq r \dfrac{\mathcal D} {\sqrt{\langle \Delta\hat Z^2 \rangle}} \label{Ntrans} ,
\end{align}
resulting in a $\sqrt{N_{\rm tr}} \sim r$ scaling.

\section{F. Dephasing channel: $N$-dependence via exact calculations}
The fidelity of initial GHZ states undergoing Markovian dephasing (exactly) obeys
\begin{equation}
{\cos^2\mathcal D} = \frac{1+e^{-N\gamma\tau}\cos N\omega_0\tau}{2} ,
\label{FDiretoGHZ}
\end{equation}
where the $\tau\sim1/N$ scaling is clear.

For the separable, symmetric state of $N$ qubits, each initially on the equator of its Bloch sphere, it is given by
\begin{equation}
{\cos^2\mathcal D} = \frac{1}{2^N}\left(1+e^{-\gamma\tau}\cos\omega_0\tau \right)^N ,
\label{FDiretoSep}
\end{equation}
and can be rewritten and expanded as
\begin{align}
e^{-\gamma\tau}\cos(\omega_0\tau) &= 2\cos^{2/N}\mathcal D-1 , \label{exactinvert}\\
\left( 1 - \gamma\tau+\frac{\gamma^2\tau^2}{2} + \mathcal{O}(\gamma^3\tau^3)  \right)       \left(  1 - \frac{\omega_0^2\tau^2}{2} + \mathcal{O}(\omega_0^4\tau^4)  \right) 
&= 1 + 4\frac{\ln\cos\mathcal D}{N} + \mathcal{O}(1/N^2) , \label{exactexpans}
\end{align}
where we make use of the fact that $\tau$ vanishes for increasing values of $N$. The above expression can be simplified to
\begin{equation}
\gamma\tau + \frac{\omega_0^2-\gamma^2}{2} \tau^2 +\mathcal{O}(\gamma^3\tau^3,\omega^3\tau^3) =  4\frac{\ln\sec\mathcal D}{N} + \mathcal{O}(1/N^2) .
\label{exactexpansfinal}
\end{equation}
The two regimes can be seen in the above equation: for large values of $N$ the first term of the left-hand side is dominant, yielding a $\tau\sim1/N$ dependence; for smaller values of $N$, the second term is dominant,  and $\tau\sim1/\sqrt{N}$. We can estimate when this transition occurs by finding the value of $\tau$ that leads to equal contributions from both terms of the left-hand side of \eqref{exactexpansfinal}, which is $\tau_{\rm tr}=2\gamma/(\omega_0^2-\gamma^2)$. The corresponding value of $N$ is
\begin{equation}
N_{\rm tr} =  (r^2 - 1) \ln \sec\mathcal D ,
\label{Nr2}
\end{equation}
so that for $r\gg1$ the transition happens around a value $N$ proportional to $r^2$.


\begin{thebibliography}{99}{
\bibitem{MandelstamTamm} L. Mandelstam and I. G. Tamm,  J. Phys. (Moscow) \textbf{9}, 249 (1945).
\bibitem{Vaidman} L. Vaidman, Am. J. Phys {\bf 60}, 182 (1992).
\bibitem{EberlySingh} J. H. Eberly and L. P. S. Singh, Phys. Rev. D {\bf 7}, 359 (1973).
\bibitem{LeubnerKiener} C. Leubner and C. Kiener, Phys. Rev. A {\bf 31}, 483 (1985).
\bibitem{Fleming} G. N. Fleming, Nuovo Cimento A {\bf 16}, 263 (1973).
\bibitem{Bhattacharyya} K. Bhattacharyya, J. Phys. A: Math. Gen. {\bf 16}, 2993 (1983).
\bibitem{GisSabelli} E. A. Gislason, N. H. Sabelli, and J. W. Wood, Phys. Rev. A {\bf 31}, 2078 (1985).
\bibitem{BauerMello} M. Bauer and P. A. Mello, Proc. Nat. Acad. Sci. USA {\bf 73}, 283 (1976);  Annals of Physics {\bf 111}, 38 (1978).
\bibitem{Uhlmann} A. Uhlmann, Phys. Lett. A {\bf 161}, 329 (1992).
\bibitem{UffHilge}J. Uffink and J. Hilgevoord, Found. Phys. {\bf 15}, 925 (1985).
\bibitem{Uffink} J. Uffink, Am. J. Phys.  {\bf 61}, 935 (1993).
\bibitem{Pfeifer} P. Pfeifer, Phys. Rev. Lett. {\bf 70}, 3365 (1993).
\bibitem{PfFrohlich} P. Pfeifer and J. Frolich, Rev. Mod. Phys. {\bf 67}, 759 (1995).
\bibitem{AA}J. Anandan and Y. Aharonov, Phys. Rev. Lett. {\bf 65}, 1697 (1990).
\bibitem{Anandan}J. Anandan, Found. Phys {\bf 21}, 1265 (1991).
\bibitem{HoreshMann} N. Horesh and A. Mann, J. Phys. A: Math. Gen. {\bf 31}, L609 (1998).
\bibitem{Pati} A. K. Pati, Phys. Lett. A {\bf 262}, 296 (1999).
\bibitem{MargolusLevitin} Norman Margolus and Lev B. Levitin, Physica D \textbf{120}, 188 (1998).
\bibitem{Soderholm} J. Soderholm, G. Bjork, T. Tsegaye, and A. Trifonov, Phys. Rev. A {\bf 59}, 1788 (1999).
\bibitem{LevitinToffoli} Lev B. Levitin and T. Toffoli, Phys. Rev. Lett. {\bf 103}, 160502 (2009).
\bibitem{GiovannettiPRA} V. Giovannetti, S. Lloyd, and L. Maccone, Phys. Rev. A {\bf 67}, 052109 (2003).
\bibitem{GiovannettiEPLJOB} V. Giovannetti, S. Lloyd, and L. Maccone, Europhys. Lett. {\bf 62}, 615 (2003); J. Opt. B {\bf 6}, S807 (2004).
\bibitem{Batle} J. Batle, M. Casas, A. Plastino, and A.R. Plastino, Phys. Rev. A {\bf 72}, 032337 (2005).
\bibitem{Borras2006} A. Borras, M. Casas, A. R. Plastino, and A. Plastino, Phys. Rev. A {\bf 74}, 022326 (2006).
\bibitem{Zander}C. Zander, A. R. Plastino, A. Plastino, and M. Casas, J. Phys. A: Math. Theor. {\bf 40}, 2861 (2007). 
\bibitem{Kupferman} J. Kupferman and B. Reznik, Phys. Rev. A {\bf 78}, 042305 (2008).
\bibitem{Frowis}F. Frowis, Phys. Rev. A {\bf 85}, 052127 (2012).
\bibitem{Andrecut} M. Andrecut and M. K. Ali, J. Phys. A {\bf 37}, L157 (2004).
\bibitem{GrayVogt} J. E. Gray and A. Vogt, J. Math. Phys. {\bf 46}, 052108 (2005).
\bibitem{LuoZhang} S. Luo and Z. Zhang, Lett. Math. Phys. {\bf 71}, 1 (2005).
\bibitem{Andrews} M. Andrews, Phys. Rev. A {\bf 75}, 062112 (2007).
\bibitem{ZielinskiZych} B. Zielinski and M. Zych, Phys. Rev. A {\bf 74}, 034301 (2006).
\bibitem{Yurtsever} U. Yurtsever, Phys. Scr. {\bf 82}, 035008 (2010).
\bibitem{FuLiLuo} S.-S. Fu, N. Li, and S. Luo, Commun. Theor. Phys. {\bf 54}, 661 (2010).
\bibitem{Chau} H. F. Chau, Phys. Rev. A {\bf 81}, 062133 (2010).
\bibitem{Deffner} S. Deffner and E. Lutz, arXiv:1104.5104 (2011).
\bibitem{Brody} D. C. Brody, J. Phys. A: Math. Theor. {\bf44} 252002 (2011).
\bibitem{Ashhab} S. Ashhab, P. C. de Groot, and F. Nori, Phys. Rev. A {\bf85}, 052327 (2012).
\bibitem{Zurek} W. H. Zurek, Rev. Mod. Phys. {\bf 75}, 715 (2003).
\bibitem{JonesKok} P. J. Jones and P. Kok, Phys. Rev. A {\bf 82}, 022107  (2010).
\bibitem{CarliniBr8} A. Carlini, A. Hosoya, T. Koike, and Y. Okudaira, J. Phys. A: Math. Theor. {\bf41}, 045303 (2008).
\bibitem{Beretta} G. P. Beretta, arXiv:quant-ph/0511091 (2006).
\bibitem{Obada} A.-S. F. Obada, D. A. M. Abo-Kahla, N. Metwally, and M. Abdel-Aty, Physica E {\bf43}, 1792 (2011).
\bibitem{BrodyN} D. C. Brody and E.-M. Graefe, arXiv:1208.5297v1
\bibitem{BraunsteinCaves} S. L. Braunstein and C. M. Caves, Phys. Rev. Lett. {\bf 72}, 3439 (1994)
\bibitem{Uhlmann2} A. Uhlmann, ``The Metric of Bures and the Geometric Phase'', in R. Gielerak, J. Lukierski, and Z. Popowicz (Eds.), Quantum Groups and Related Topics: Proceedings of the First Max Born Symposium, 267, 1992.
\bibitem{Supp} See Supplemental Material below for the derivation of the general bound, minimization of the quantum Fisher information for dephasing, as well as details of the $N$ dependence of the bound for the $N$-qubit system.
\bibitem{arccos} $\arccos$ defined on $[0,\pi]$ throughout the article.
\bibitem{Boixo} S. Boixo, S. T. Flammia, C. M. Caves, and J. M. Geremia, Phys. Rev. Lett. {\bf 98}, 090401 (2007).
\bibitem{BLNR} B. M. Escher, L. Davidovich, N. Zagury, and R. L. de Matos Filho, Phys. Rev. Lett. {\bf 109}, 190404 (2012).
\bibitem{BRL} B. M. Escher, R. L. de Matos Filho, and L. Davidovich, Nature Phys. {\bf 7}, 406 (2011);  Braz. J. Phys. {\bf 41}, 229, (2011).
}
\end{thebibliography}
\end{document}